\documentclass[12pt]{iopart}
\usepackage{iopams}
\usepackage{graphicx}
\usepackage{setstack}
\usepackage{stackrel}
\usepackage{color}
\definecolor{darkblue}{rgb}{0,0,0.6}
\definecolor{darkred}{rgb}{0.6,0,0}
\definecolor{darkgreen}{rgb}{0,0.6,0}
\usepackage[colorlinks=true,urlcolor=darkblue,citecolor=darkblue,linkcolor=darkred,hyperfootnotes=false]{hyperref}

\begin{document}
\title{Tradeoff of generalization error in unsupervised learning}


\author{Gilhan Kim$^1$, Hojun Lee$^1$, Junghyo Jo$^2$, and Yongjoo Baek$^1$}

\address{$^1$ Department of Physics and Astronomy \& Center for Theoretical Physics, Seoul National University, Seoul 08826, Korea}
\address{$^2$ Department of Physics Education \& Center for Theoretical Physics, Seoul National University, Seoul 08826, Korea}
\ead{y.baek@snu.ac.kr}

\date{\today}

\begin{abstract}
Finding the optimal model complexity that minimizes the generalization error (GE) is a key issue of machine learning. For the conventional supervised learning, this task typically involves the bias-variance tradeoff: lowering the bias by making the model more complex entails an increase in the variance. Meanwhile, little has been studied about whether the same tradeoff exists for unsupervised learning. In this study, we propose that unsupervised learning generally exhibits a two-component tradeoff of the GE, namely the model error and the data error---using a more complex model reduces the model error at the cost of the data error, with the data error playing a more significant role for a smaller training dataset. This is corroborated by training the restricted Boltzmann machine to generate the configurations of the two-dimensional Ising model at a given temperature and the totally asymmetric simple exclusion process with given entry and exit rates. Our results also indicate that the optimal model tends to be more complex when the data to be learned are more complex.
\end{abstract}

\maketitle

\section{Introduction}
Inductive reasoning, which derives general principles from specific observations, is an essential generalization process that builds up human knowledge. With the advent of big data, there is a rapidly growing need for automating the process of inductive reasoning. Lately, much development has been made in this direction thanks to various machine learning techniques based on the artificial neural networks (ANNs)~\cite{goodfellow2016deep,mehta2019high,carleo2019machine,bahri2020statistical}. However, despite their tremendous success, we still have a very limited understanding of when and how an ANN achieves a good generalization capacity.

There are two major types of generalization tasks performed by the ANNs. The more well-studied type is {\em supervised learning}, which refers to the task of guessing the correct form of a function over its entire domain by generalizing some given examples of its functional relations. In this case, the failure to properly generalize the given examples, the generalization error (GE), can be defined in terms of the mean squared error (MSE) of the predicted function from the true function. In practice, the true function is unknown, so the MSE estimated using independently drawn examples of functional relations (called the {\em test error}) is used as a proxy for the GE.

Thanks to the mathematical structure of the MSE, this GE is readily decomposed into two parts~\cite{mehta2019high}. The first part, called the bias, quantifies how the predicted function on average deviates from the true function. The second part, called the variance, quantifies how much the predicted function fluctuates from its average behavior. In many examples of supervised learning, these two components of the GE exhibit a tradeoff behavior: as the model complexity ({\em e.g.}, the size of the ANN) increases, the bias decreases at the cost of the increasing variance. This leads to the GE showing a U-shape dependence on the model complexity, which is called the {\em bias-variance tradeoff}. It should be noted that the decomposition is not limited to the MSE but also generalized to different types of the GE (see, for example, \cite{kohavi1996bias}). In addition, according to recent studies, the GE of supervised learning again exhibits a monotonic decrease if the model complexity is increased even further. This is called the {\em double descent phenomenon}, whose origin has been extensively discussed~\cite{belkin2019reconciling,spigler2019jamming,nakkiran2021deep, rocks2022memorizing}.

Meanwhile, there is the less studied but no less important type of tasks, namely {\em unsupervised learning}, which refers to the task of finding the probability distribution that best captures the statistical properties of a dataset sampled from an unknown distribution. The GE for unsupervised learning can be defined as the Kullback-Leibler (KL) divergence of the predicted distribution from the true distribution to be found. Again, there has been a proposal about how this GE can be decomposed into the bias and the variance~\cite{heskes1998bias}. However, little has been studied about whether the GE of unsupervised learning also exhibits a tradeoff behavior.

In this study, we address the problem by training the Restricted Boltzmann Machine (RBM) to learn the data generated from the two-dimensional (2-d) Ising model and the totally asymmetric simple exclusion process (TASEP), which are well-known models of equilibrium and nonequilibrium phase transitions, respectively. Since the distributions of the configurations of these models are exactly known, the GE can be calculated exactly. Examining how these quantities depend on the number of hidden nodes of the RBM, we observe that the GE exhibits a tradeoff behavior. We propose a two-component decomposition scheme of the GE so that the tradeoff is determined by the monotonic behaviors of the two error components, one related to the model limitations and the other to the fluctuations in the data. We also examine how the optimal model complexity, resulting from by a tradeoff among these three GE components, depends on the complexity of the given training dataset.

The rest of the paper is organized as follows. In Sec.~\ref{sec:theory}, we introduce the restricted Boltzmann machine and our decomposition scheme for its generalization error. In Sec.~\ref{sec:methods}, we describe how the RBM is trained using the data generated from the 2-d Ising model and the TASEP. In Sec.~\ref{sec:results}, we present our results about the tradeoff behaviors observed in unsupervised learning of the RBMs. Finally, we summarize our findings and discuss future works in Sec.~\ref{sec:summary}.

\begin{figure}
\includegraphics[width=\textwidth]{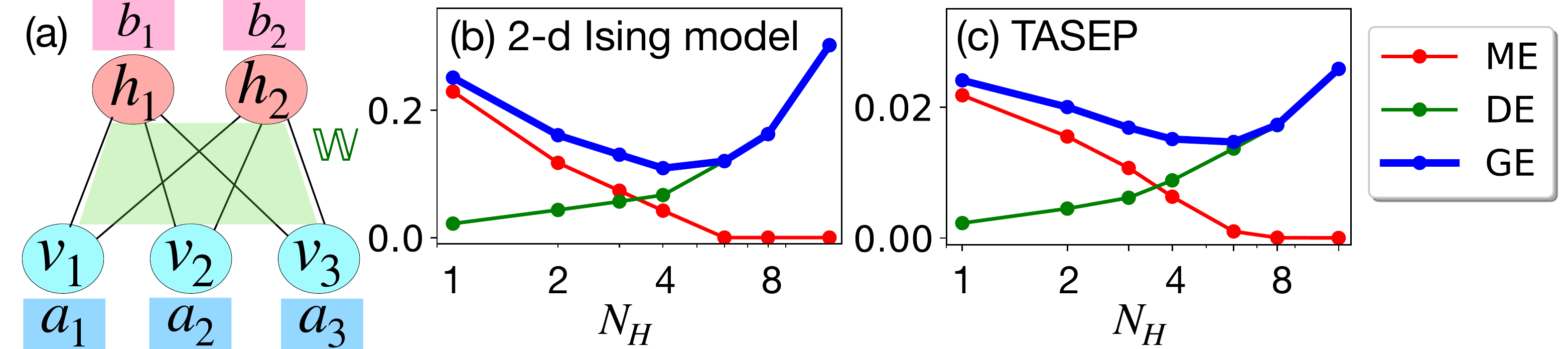}
\caption{\label{fig:fig1} (a) A schematic diagram showing the structure of the RBM. The generalization error (GE) of the RBM depends nonmonotonically on the number of hidden nodes $N_H$, exhibiting a tradeoff between the data error (DE) and the model error (ME) as illustrated by the generation tasks of (b) the 2-d Ising configurations at temperature $T = 3.6$ and (c) the TASEP configurations at $\alpha = \beta = 0.7$. The lines are to guide the eye.}
\end{figure}

\section{Theory} \label{sec:theory}

\subsection{Restricted Boltzmann Machine}

The RBM is an energy-based generative model proposed by Smolensky~\cite{smolensky1986information} and popularized by Hinton~\cite{hinton2002neural,hinton2012practical}. It has the form of a bipartite network of $N_V$ nodes in the visible layer and $N_H$ nodes in the hidden layer, see Fig.~\ref{fig:fig1}(a). For the visible-layer configuration $\mathbf{v} \equiv \{v_i\}_{i=1}^{N_V}$ and the hidden-layer configuration $\mathbf{h} \equiv \{h_i\}_{i=1}^{N_H}$, the corresponding energy is given by
\begin{eqnarray} \label{eq:energy}
	E(\mathbf{v},\mathbf{h})=-\mathbf{a}^\mathrm{T}\,\mathbf{v}-\mathbf{b}^\mathrm{T}\,\mathbf{h}-\mathbf{v}^\mathrm{T}\,\mathbb{W}\,\mathbf{h},
\end{eqnarray}
where $\mathbf{a} \equiv \{a_i\}_{i=1}^{N_V}$ and $\mathbf{b} \equiv \{b_i\}_{i=1}^{N_H}$ indicate the bias terms, and $\mathbb{W}$ is the $N_V \times N_H$ weight matrix coupling the two layers. The state of each node is a Boolean variable, {\em i.e.}, $v_i \in \{0,\,1\}$ and $h_i \in \{0,\,1\}$. The probability of each configuration is determined by this energy function according to the Boltzmann distribution
\begin{eqnarray} \label{eq:joint_dist}
	Q_{VH}(\mathbf{v},\mathbf{h})=\frac{1}{Z}\exp\left[-E(\mathbf{v},\mathbf{h})\right],
\end{eqnarray}
where $Z$ is the normalizing factor (or the {\em partition function})
\begin{eqnarray} \label{eq:partition_function}
	Z \equiv \sum_{\mathbf{v},\mathbf{h}}\exp\left[-E(\mathbf{v},\mathbf{h})\right].
\end{eqnarray}
The goal of the RBM is to find $\mathbf{a}$, $\mathbf{b}$, and $\mathbb{W}$ such that the marginal distribution $Q_V(\mathbf{v}) \equiv \sum_\mathbf{h} Q_{VH}(\mathbf{v},\mathbf{h})$ is as similar as possible to some given empirical distribution $P_V(\mathbf{v})$. To put it precisely, the RBM seeks to achieve
\begin{eqnarray}
	Q^*_V \equiv \;\stackrel[Q_V]{}{\mathrm{arg\,min}} D_\mathrm{KL}(P_V\|Q_V),
\end{eqnarray}
where
\begin{eqnarray} \label{eq:KLD}
D_\mathrm{KL}(P_V\|Q_V) \equiv \sum_{\mathbf{v}}P_V(\mathbf{v})\log\frac{P_V(\mathbf{v})}{Q_V(\mathbf{v})}
\end{eqnarray}
is the Kullback-Leibler (KL) divergence. Towards this purpose, the above KL divergence is taken to be the loss function, and the RBM is updated according to the gradient descent
\begin{eqnarray}
	a_{i}(t+1) &= a_{i}(t) -\alpha\frac{\partial}{\partial a_{i}}D_\mathrm{KL}(P_V\|Q_V), \label{eq:gd_a}\\
	b_{i}(t+1) &= b_{i}(t) -\alpha\frac{\partial}{\partial b_{i}}D_\mathrm{KL}(P_V\|Q_V), \label{eq:gd_b}\\
	W_{ij}(t+1) &= W_{ij}(t) -\alpha\frac{\partial}{\partial W_{ij}}D_\mathrm{KL}(P_V\|Q_V). \label{eq:gd_w}
\end{eqnarray}	
Denoting by
\begin{eqnarray}
	Q_{H|V}(\mathbf{h}|\mathbf{v}) \equiv \frac{Q_{VH}(\mathbf{v},\mathbf{h})}{Q_V(\mathbf{v})}
	= \prod_{j=1}^{N_H} \frac{\exp\left[b_j h_j+\sum_{i=1}^{N_V}v_i W_{ij} h_j\right]}{1+\exp\left[b_j+\sum_{i=1}^{N_V}v_i W_{ij}\right]}
\end{eqnarray}
the conditional probability of the hidden-layer configuration, we can show that the gradients of the KL divergence satisfy
\begin{eqnarray}
	\frac{\partial}{\partial a_{i}}D_\mathrm{KL}(P_V\|Q_V) &= \langle v_i \rangle_{Q_V}-\langle v_i \rangle_{P_V}, \label{eq:a_grad}\\
	\frac{\partial}{\partial b_{i}}D_\mathrm{KL}(P_V\|Q_V) &= \langle h_i \rangle_{Q_{VH}}-\langle h_i \rangle_{P_V Q_{H|V}}, \label{eq:b_grad}\\
	\frac{\partial}{\partial W_{ij}}D_\mathrm{KL}(P_V\|Q_V) &= \langle v_i h_j \rangle_{Q_{VH}}-\langle v_i h_j \rangle_{P_V Q_{H|V}}, \label{eq:w_grad}
\end{eqnarray}
where $\langle\cdot\rangle_F$ denotes an average with respect to the probability distribution $F$. Thus, the training saturates when the first and the second moments of the state variables are equal for both empirical ($P_V Q_{H|V}$) and model ($Q_{VH}$) distributions.

Since these gradients involve the average $\langle\cdot\rangle_{Q_{VH}}$ whose computation is difficult for large networks, various approximation methods are used in practice, such as contrastive divergence (CD)~\cite{hinton2002neural}, persistent contrastive divergence (PCD)~\cite{tieleman2008training}, fast PCD~\cite{tieleman2009using}, Bennett's acceptance ratio method~\cite{krause2020algorithms}, and the Thouless-Anderson-Palmer equations~\cite{gabrie2015training}. However, to avoid further complications arising from the approximations, in this study we stick to the exact gradients written above.

\subsection{Error components in unsupervised learning}

The goal of unsupervised learning is to construct a generative model whose statistical properties are as similar as possible to the true distribution underlying an ensemble of objects. We may formally describe the problem as follows. Suppose that there exists the true probability distribution $P^0_V$ generating an ensemble of objects. Then we can think of the best model $Q^0_V$ that the RBM can express, namely
\begin{equation}
	Q^0_V \equiv \;\stackrel[Q_V]{}{\mathrm{arg\,min}} D_\mathrm{KL}(P^0_V\|Q_V).
\end{equation}
Ideally, the RBM should generate $Q^0_V$ at the end of the training. However, this may not be true due to three reasons. First, the KL divergence is generally not a convex function of $\mathbf{a}$, $\mathbf{b}$, and $\mathbb{W}$, so the gradient descent shown in Eqs.~\eref{eq:gd_a}, \eref{eq:gd_b}, and \eref{eq:gd_w} may end up in a local minimum far away from $Q^0_V$. Second, even if the RBM does evolve towards $Q^0_V$, the convergence may take an extremely long time. In this case, the training will have to end before $Q^0_V$ is reached. Third, in practical situations, we do not have direct access to the true distribution (which we denote by $P^0_V$) generating the observed samples. Due to the sampling error, the distribution $P_V$ used in the training is generally different from $P^0_V$. Then, even if the RBM does manage to find a distribution most similar to $P_V$, it may still be quite different from $Q^0_V$.

For these reasons, the distribution generated by the RBM at the end of the training is not $Q^0_V$, but each training will result in its own model distribution $Q^\mathrm{m}_V$. Then the GE, which quantifies the expected difference of the model distribution from the true distribution, can be defined as
\begin{equation} \label{eq:ge_def}
	\mathrm{GE} \equiv \left\langle D_\mathrm{KL}(P^0_V\|Q^\mathrm{m}_V) \right\rangle_\mathrm{m},
\end{equation}
where $\langle\cdot\rangle_\mathrm{m}$ represents the average with respect to different models obtained by independent trainings. By definition, the GE cannot be smaller than the {\em model error} (ME) 
\begin{equation} \label{eq:me_def}
	\mathrm{ME} \equiv D_\mathrm{KL}(P^0_V\|Q^0_V),
\end{equation}
which indicates the fundamental lower bound on how similar the model distribution generated by the RBM can be to the true distribution $P^0_V$. Finally, we identify the excess part of the error,
\begin{equation} \label{eq:de_def}
	\mathrm{DE} \equiv \mathrm{GE} - \mathrm{ME},
\end{equation}
as the {\em data error} (DE), which mainly stems from deviations of the training data from the true distribution, as will be shown below. Thus, Eqs.~\eref{eq:ge_def}, \eref{eq:me_def}, and \eref{eq:de_def} define a two-component decomposition of the GE for unsupervised learning.

\section{Methods} \label{sec:methods}

To examine the behaviors of the error components defined above, we train the RBMs to two basic models of statistical physics: the 2-d Ising model and the TASEP with open boundaries. We chose these models for two advantages. First, these models have exactly known steady-state distributions. The Ising model follows the standard Boltzmann statistics, and the nonequilibrium steady-state statistics of the TASEP can be exactly obtained using the matrix-product ansatz~\cite{derrida1993exact}. Thus, for these models, we have full information about $P^0_V$. Second, these models provide examples of equilibrium and nonequilibrium phase transitions. Depending on the nature of the phases and the transitions between them, this allows us to control the complexity of the data and examine how the tradeoff behavior is affected by it. With these considerations in mind, we describe the two models in the following.

\begin{figure}
\includegraphics[width=\textwidth]{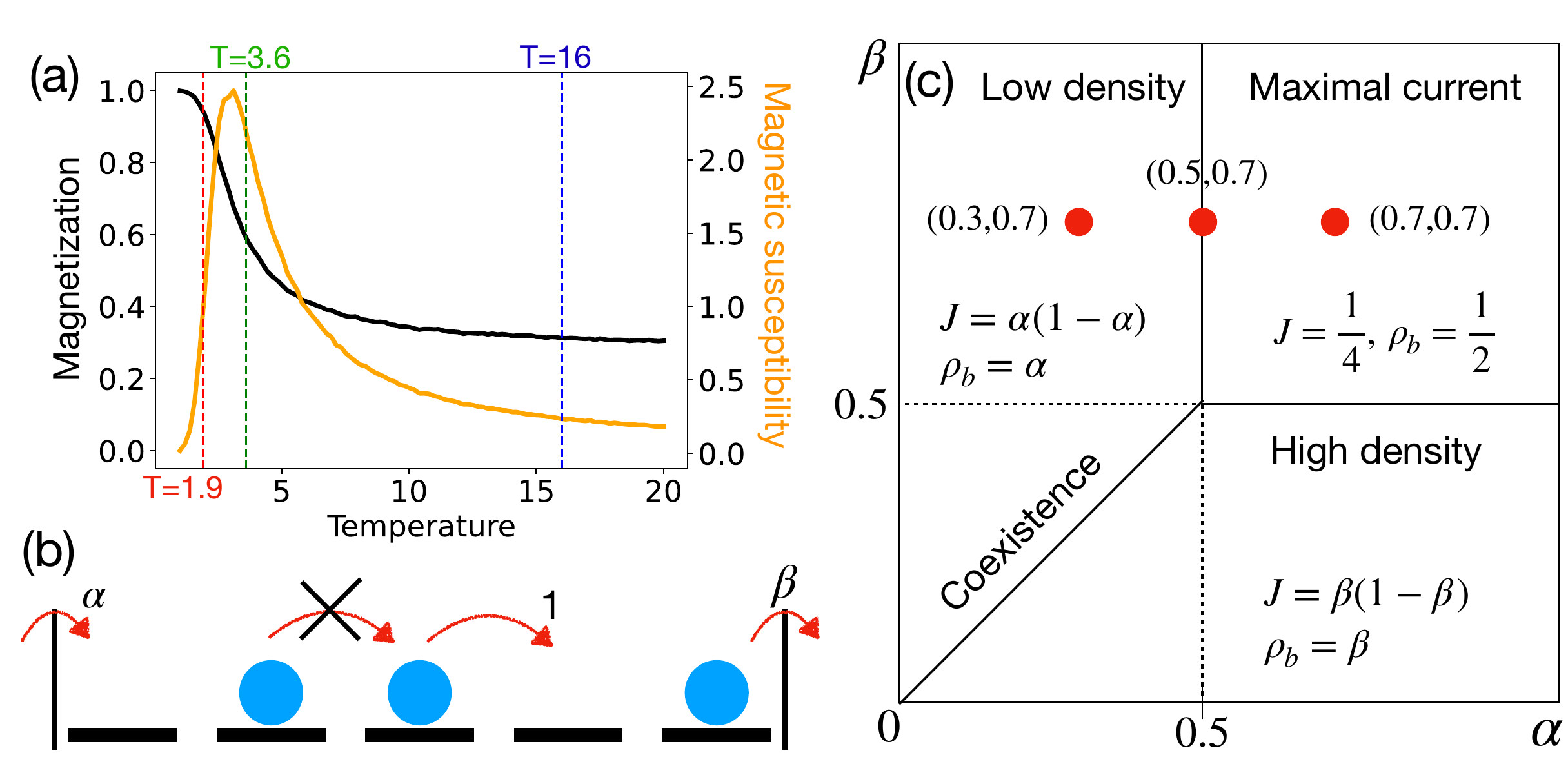}
\caption{\label{fig:training_data} (a) The magnetization and the susceptibility of the 2-d Ising model on the $3 \times 3$ square lattice with periodic boundaries. The vertical lines indicate the values of temperature used for generating the training datasets. (b) A schematic illustration of the TASEP with open boundaries. (c) The phase diagram of the TASEP with open boundaries. The red points indicate the control parameters used for generating the training datasets.}
\end{figure}

\subsection{2-d Ising model}
The Ising model, originally proposed as a simplified model of ferromagnetism~\cite{ising1925beitrag}, is a paradigmatic model of equilibrium phase transitions with exact solutions for its free energy~\cite{onsager1944crystal} and spontaneous magnetization~\cite{yang1952spontaneous} on the 2-d square lattice. In this study, we use the Ising model on a $3 \times 3$ square lattice with periodic boundary conditions ($v_{4,j}=v_{1,j}$ and $v_{i,4}=v_{i,1}$) following the Boltzmann statistics
\begin{equation}
	P^0_V(\mathbf{v}) \propto \exp\!\left[-\frac{1}{T}\mathcal{H}(\mathbf{v})\right],
\end{equation}
where the Hamiltonian $\mathcal{H}$ is given by
\begin{equation}
	\mathcal{H}(\mathbf{v}) = -\sum_{i=1}^3\sum_{j=1}^3 \left[(2v_{i,j}-1) (2v_{i+1,j}-1)+(2v_{i,j}-1) (2v_{i,j+1}-1)\right],
\end{equation}
with $v_{i,j} \in \{0,\,1\}$ for each $i$ and $j$. We train the RBMs to the equilibrium distributions at three different temperatures $T = 1.9$, $T = 3.6$, and $T = 16$. These values are chosen so that $T = 1.9$ corresponds to the ferromagnetic (ordered) phase, $T = 3.6$ stands for the critical regime, and $T = 16$ generates the paramagnetic (disordered) phase. Even though the size of the system used in our study is small, these three values of temperature generate markedly different ensembles of spin configurations, as shown in the order parameter (magnetization) and its fluctuations (susceptibility) plotted in Fig.~\ref{fig:training_data}(a).

Using $P^0_V$ thus defined, we train the RBM in two different ways. The first way is the usual unsupervised learning. We draw a certain number of equilibrium spin configurations from $P^0_V$ by the Metropolis-Hastings algorithm. To remove correlations between different sampled configurations, we saved the snapshot of the system every $90$ or more (varied depending on the correlation time) Monte Carlo steps. This sampled set of Ising configurations form $P_V$, which we use in the gradient descent dynamics described by Eqs.~\eref{eq:gd_a}--\eref{eq:w_grad}. We repeat the process $10$ times to obtain $10$ independent models $\{ Q^\mathrm{m}_V \}$, with which we calculate the GE according to Eq.~\eref{eq:ge_def}.

Meanwhile, to calculate the ME and the DE, we train the RBM in a different way. Instead of the sampled $P_V$, we use the true distribution $P^0_V$ to directly calculate the true gradients in Eqs.~\eref{eq:gd_a}--\eref{eq:w_grad}. This is done by evaluating the averages over all the $2^9 = 512$ spin configurations of the 2-d Ising model on the $3 \times 3$ square lattice. Repeating this training $10$ times, we choose the resulting model that minimizes the KL divergence to be an estimate for $Q^0_V$. In fact, we find that the KL divergence obtained by this training method exhibits only small variabilities: the error of the estimated ME is at most about the order of the symbol size in the plots.

\subsection{TASEP with open boundaries}
The TASEP is a simple model of nonequilibrium 1-d transport of hard-core particles. Originally proposed as a model of biopolymerization~\cite{macdonald1968kinetics}, the model has found numerous applications in various traffic~\cite{helbing2001traffic} and biological~\cite{chowdhury2005physics} systems.

For the case of open boundaries, the process is defined as follows (see Fig.~\ref{fig:training_data}(b) for a schematic illustration). Consider a 1-d chain of $L$ discrete sites. Each site can hold at most a single particle. From the $1$-st site to the ($L-1$)-th site, every particle moves to the right with rate $1$ if the next site is empty, and the movement is forbidden if the next site is occupied. Meanwhile, if the $1$-st site is empty ($L$-th site is occupied), a particle moves into the site from the left particle reservoir (moves from the site to the right particle reservoir) with entry rate $\alpha$ (exit rate $\beta$).

Depending on the values of $\alpha$ and $\beta$, the TASEP exhibits various phases distinguished from each other by the current $J$ and the bulk density $\rho_b$, as shown in Fig.~\ref{fig:training_data}(c). The phase boundaries can be correctly predicted by the simple kinematic wave theory or the mean-field arguments, although the exact solution for the steady-state statistics requires the matrix product ansatz~\cite{derrida1993exact,blythe2007nonequilibrium}.

Similarly to the case of the 2-d Ising model, we train the RBM in two different ways. First, we generate a certain number of steady-state particle configurations by directly running the kinetic Monte Carlo simulation of the TASEP. To remove correlations between different sampled configurations, we take the snapshot every $90$ Monte Carlo steps. This sampled set is used as $P_V$ in the gradient descent dynamics described by Eqs.~\eref{eq:gd_a}--\eref{eq:w_grad}. Then the same process is repeated $10$ times to obtain $10$ independent models $\{ Q^\mathrm{m}_V \}$, with which we calculate the GE according to Eq.~\eref{eq:ge_def}.

To calculate the ME, we need the true distribution $P^0_V$ of the TASEP. Using the recursive relations between matrix-product representations of the steady-state probabilities shown in \cite{blythe2007nonequilibrium}, we can iteratively determine the probabilities of $2^9 = 512$ TASEP configurations. Then, based on these probabilities, we directly calculate the true gradients in Eqs.~\eref{eq:gd_a}--\eref{eq:w_grad}. Then the ME is found by choosing the model that minimizes the KL divergence to be an estimate for $Q^0_V$.

The codes for the training methods described above are available in \cite{github}.

\begin{figure}
\includegraphics[width=\columnwidth]{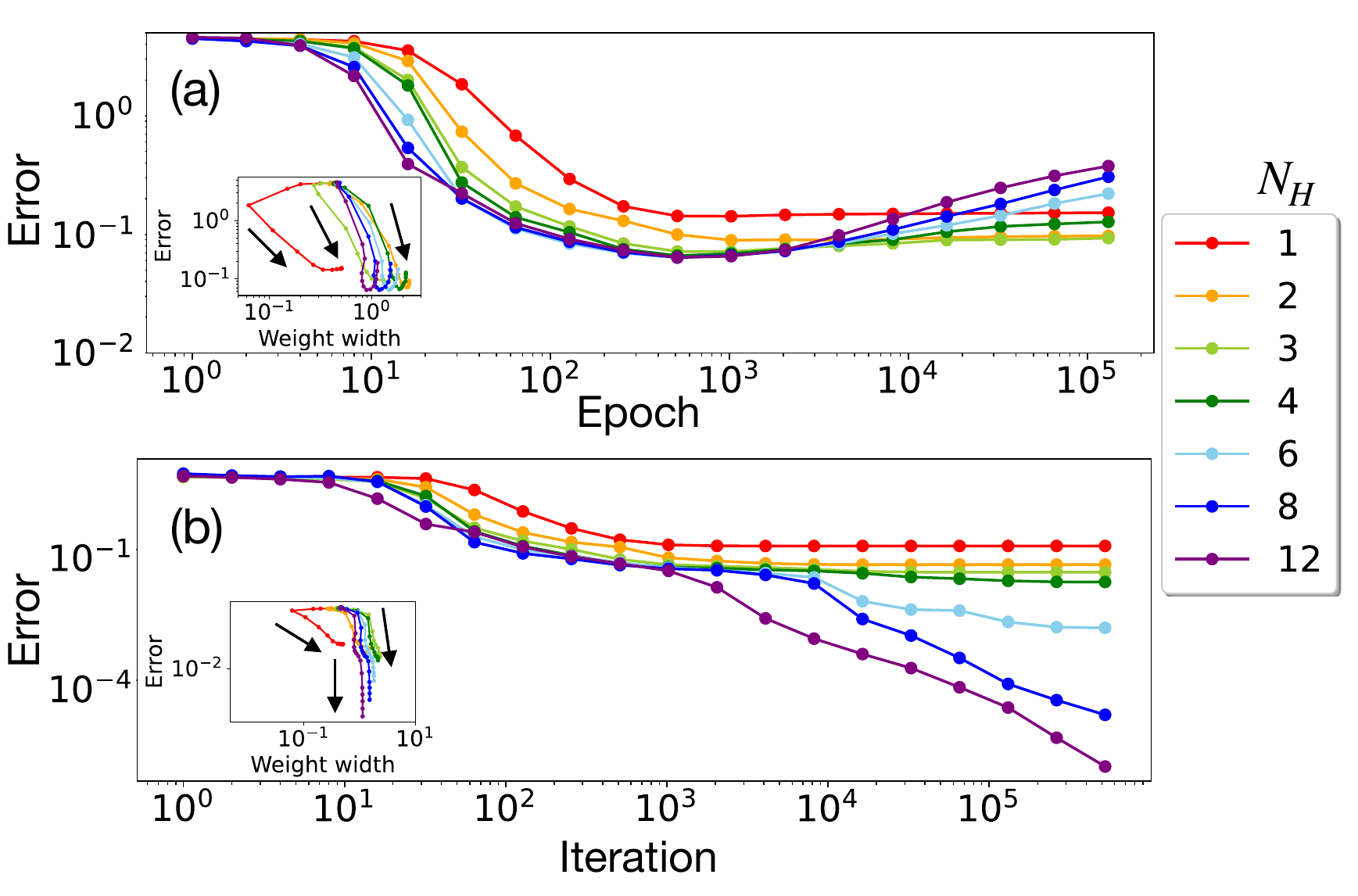}
\caption{\label{fig:dynamics} Training dynamics of the RBM. (a) The evolution of the generalization error (GE) when the RBM is trained to $512$ sampled configurations of the 2-d Ising model at temperature $T = 1.9$. (b) The evolution of the GE when the RBM is trained to the exact gradients of the KL divergence at the same temperature. The insets show the average trajectories of the RBM in terms of the error and the weight width during each dynamics, with the direction of time shown by black arrows. The error bars are comparable to the symbol size.}
\end{figure}

\section{Results} \label{sec:results}

\subsection{Training dynamics}

In Fig.~\ref{fig:dynamics}(a), we show the evolution of the mean error $\left\langle D_\mathrm{KL}(P^0_V\|Q^m_V) \right\rangle_m$ as the RBM is trained to the $512$ sampled configurations (using mini-batches of size $256$) of the 2-d Ising model at $T = 1.9$ (ferromagnetic phase). When $N_H \le 2$, the error monotonically decreases as the training proceeds, saturating by epoch $10^3$. However, for $N_H \ge 3$, the error reaches the minimum around epoch $500$ and increases again, reaching higher value as $N_H$ is increased. These tendencies reflect that more complex RBMs (with larger $N_H$) end up learning even the noisy features of the sampled configurations that deviate from the true $P^0_V$. This is equivalent to what we call the overfitting phenomenon in supervised learning.

Meanwhile, in Fig.~\ref{fig:dynamics}(b), we show the evolution of the same mean error as the RBM is trained to the true distribution of the 2-d Ising model at $T = 1.9$. When $N_H \le 4$, the error saturates to some value that decreases as $N_H$ is increased. When $N_H$ is increased further, the error keeps decreasing as the training proceeds, never saturating within the observation time span. These show that the overfitting effect is absent when the true distribution is directly used for the training.

The presence of overfitting can also be checked by observing the weights of the RBM. As shown in the inset of Fig.~\ref{fig:dynamics}(a), when the RBM is trained to the sampled dataset, the width (standard deviation) of the weight distribution tends to increase towards the end of the training. This is because here the RBM is effectively decreasing its temperature, trying to distinguish configurations which happen to appear in the sampled dataset from those which happen not to appear, even if they are very similar to each other. In contrast, when the RBM is trained to the true distribution, the weight width saturates to some value and stops increasing, as shown in Fig.~\ref{fig:dynamics}(b). This shows that there is no overfitting involved in the dynamics.

\subsection{Effects of data volume}

\begin{figure}
\includegraphics[width=\columnwidth]{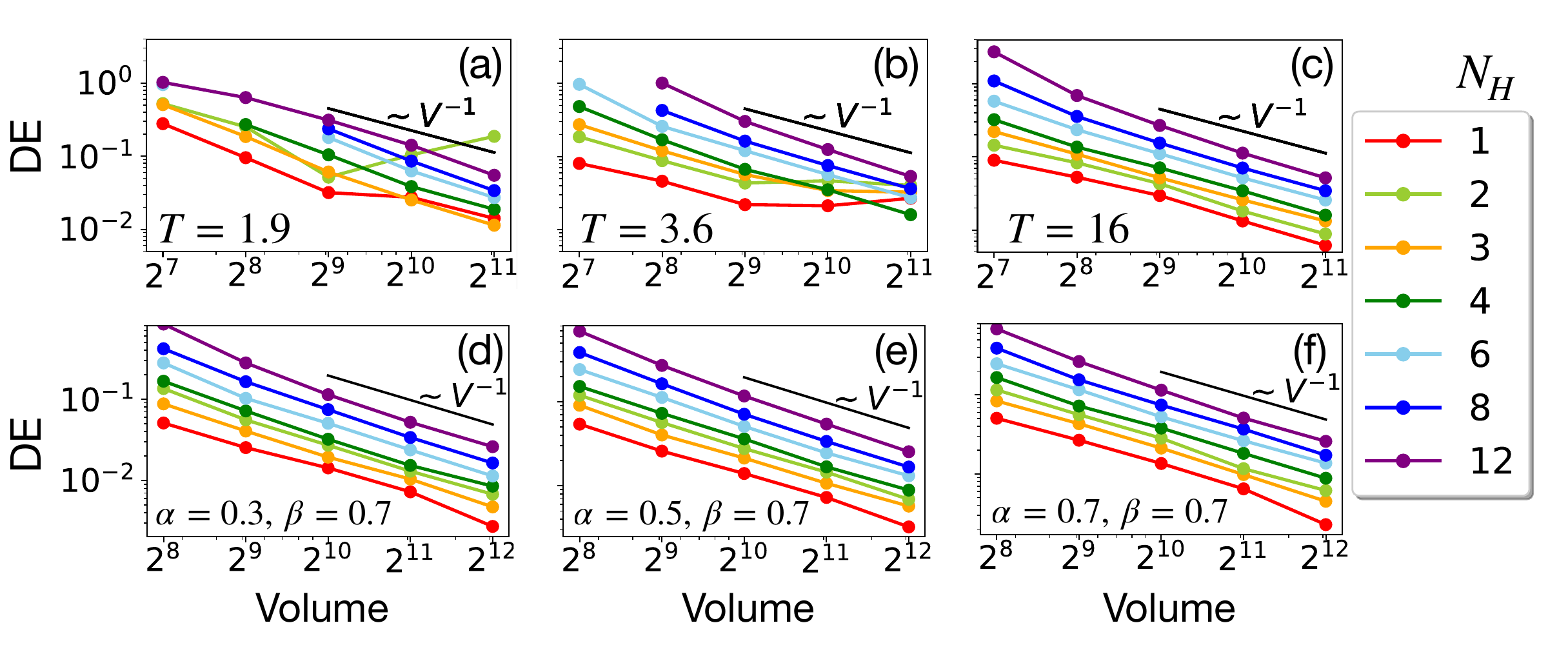}
\caption{\label{fig:de_vol} Dependence of the data error (DE) on the training dataset volume. We show results for the 2-d Ising model at (a) $T = 1.9$ (ferromagnetic), (b) $T = 3.6$ (critical), (c) $T = 16$ (paramagnetic) and for the TASEP with open boundaries at (d) $\alpha = 0.3$, $\beta = 0.7$ (low density), (e) $\alpha = 0.5$, $\beta = 0.7$ (phase boundary), (f) $\alpha = 0.7$, $\beta = 0.7$ (maximal current). The error bars are comparable to the symbol size.}
\end{figure}

As stated in Eq.~\eref{eq:me_def}, the ME depends only on the true distribution $P^0_V$ and the optimal model distribution $Q^0_V$; thus the ME does not depend on the volume of the training dataset. The only part of the GE that can be affected by the data volume is the DE, whose behaviors are shown in Fig.~\ref{fig:de_vol}. As the data volume $V$ increases, the DE tends to decrease like $V^{-1}$. This scaling behavior can be understood as follows. By careful sampling, all $V$ samples of the training dataset can be made independently and identically distributed. We note that $P_V(\mathbf{v})$ can be interpreted as the sample mean of a random variable whose value is $1$ when the sample is in the state $\mathbf{v}$ and zero otherwise. Then, when $V$ is very large, the central limit theorem implies that the sample mean $P_V(\mathbf{v})$ deviates from the true probability $P_V^0(\mathbf{v})$ by an amount proportional to $V^{-1/2}$. Then $|Q_V^m(\mathbf{v})-Q_V(\mathbf{v})|$, with $Q^0_V$ related to $P^0_V$ and $Q^\mathrm{m}_V$ to $P_V$, would also be of order $V^{-1/2}$ for every $\mathbf{v}$. Now, the DE defined in Eq.~\eref{eq:de_def} can be rewritten as
\begin{equation}
\mathrm{DE} = \left\langle D_\mathrm{KL}(P^0_V\|Q^\mathrm{m}_V) \right\rangle_\mathrm{m} - D_\mathrm{KL}(P^0_V\|Q^0_V),
\end{equation}
which reaches the minimum zero when $Q^0_V$ and $Q^\mathrm{m}_V$ are exactly equal. However, as discussed above, we expect these distributions to differ by a small amount proportional to $V^{-1/2}$ for every configuration. Since the DE is close to its minimum, this difference of order $V^{-1/2}$ leads to a correction to the DE of order $(V^{-1/2})^2 \sim V^{-1}$. Hence, the DE scales like $V^{-1}$.

We note that some deviations from $\mathrm{DE} \sim V^{-1}$ are observed for the RBMs with $N_H = 1$ and $2$ trained to the Ising model at $T = 1.9$ and $T = 3.6$. When $N_H$ is too small compared to the complexity of the training dataset, the dynamics of the RBM may develop glassy features (as is the case for supervised learning~\cite{baityjesi2019comparing}), such as the rugged loss function landscape and the presence of multiple local minima. In such cases, the DE may not be entirely induced by the difference between $P^0_V$ and $P^m_V$. However, in the regime of large $N_H$ where the DE becomes a significant part of the GE, we expect the DE to be dominated by the effects of $P^0_V \neq P^m_V$.

\subsection{Effects of $N_H$}

\begin{figure}
\includegraphics[width=\textwidth]{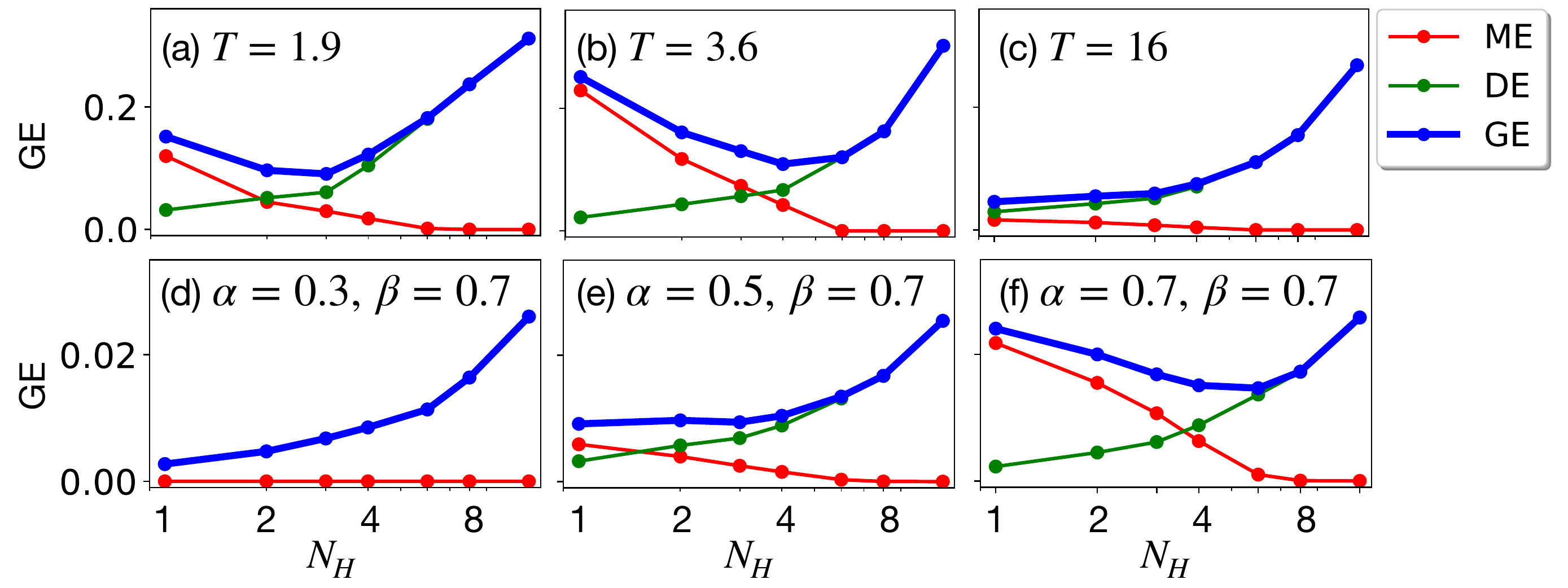}
\caption{\label{fig:tradeoff} Dependence of the generalization error (GE), the data error (DE), and the model error (ME) on the number of hidden nodes in the RBM. We show results for the 2-d Ising model at (a) $T = 1.9$ (ferromagnetic), (b) $T = 3.6$ (critical), (c) $T = 16$ (paramagnetic) and for the TASEP with open boundaries at (d) $\alpha = 0.3$, $\beta = 0.7$ (low density), (e) $\alpha = 0.5$, $\beta = 0.7$ (phase boundary), (f) $\alpha = 0.7$, $\beta = 0.7$ (maximal current). The error bars are comparable to the symbol size.}
\end{figure}

Now we examine the effects of $N_H$ on the the GE and its two components defined by Eqs.~\eref{eq:ge_def}, \eref{eq:me_def}, \eref{eq:de_def}, which are shown in Fig.~\ref{fig:tradeoff}. The results for the 2-d Ising model are shown in Fig.~\ref{fig:tradeoff}(a), (b), and (c). In all cases, the ME (DE) monotonically decreases (increases) with $N_H$. For $T = 1.9$ and $T = 3.6$, this leads to the nonmonotonic behavior of the GE, which is minimized at $N_H = 3$ for $T = 1.9$ and $N_H = 4$ for $T = 3.6$. Meanwhile, for $T = 16$, the ME is already very small for $N_H = 1$, which reflects that a single hidden node is enough to describe the disordered state, where all spins are unbiased i.i.d. random variables.

The results for the TASEP are shown in the lower panel of Fig.~\ref{fig:tradeoff}. For $\alpha = 0.3$ and $\beta = 0.7$, the ME is almost always zero, see Fig.~\ref{fig:tradeoff}(d). This is because, in the low-density phase, the occupancies of the most of the sites are i.i.d. random variables. The monotonic decrease of the ME becomes more visible as the system crosses the phase boundary (see Fig.~\ref{fig:tradeoff}(e)), although the tradeoff behavior of the GE becomes clear only when the system is well within the maximal-current phase, see Fig.~\ref{fig:tradeoff}(f). For the TASEP, it is known that the mean-field approximations are valid sufficiently far away from the boundaries. While those boundary effects decay exponentially with distance in the low-density and the high-density phases, the decay is algebraic in the maximal-current phase. Thus, the amount of correlations present in the data tend to be larger for the maximal-current phase.

To sum up, the GE exhibits a U-shaped tradeoff behavior as the ME (DE) decreases (increases) monotononically with $N_H$. The GE minimum tends to occur at a higher $N_H$ when the dataset to be modeled is more complex. The situation is analogous to the bias-variance tradeoff observed in supervised learning, as detailed in Table~\ref{tab}.


\subsection{Comparison with the bias-variance decomposition}

\begin{figure}
\includegraphics[width=\columnwidth]{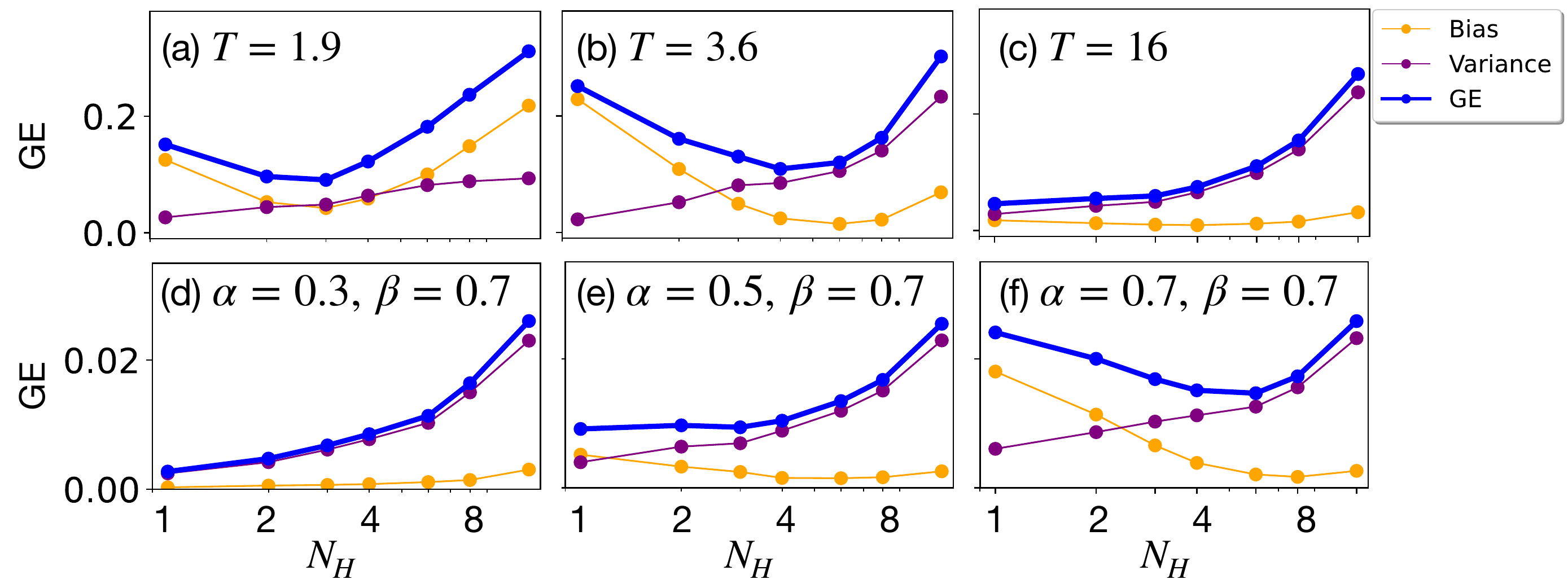}
\caption{\label{fig:heskes} Decomposition of the generalization error (GE) according to the bias-variance scheme \`{a} la Heskes. We show results for the 2-d Ising model at (a) $T = 1.9$ (ferromagnetic), (b) $T = 3.6$ (critical), (c) $T = 16$ (paramagnetic) and for the TASEP with open boundaries at (d) $\alpha = 0.3$, $\beta = 0.7$ (low density), (e) $\alpha = 0.5$, $\beta = 0.7$ (phase boundary), (f) $\alpha = 0.7$, $\beta = 0.7$ (maximal current). The error bars are comparable to the symbol size.}
\end{figure}

While the behaviors of the ME and the DE shown in Fig.~\ref{fig:tradeoff} are similar to those shown by the bias and the variance in supervised learning, there are no clear mathematical connections between these quantities. In fact, in 1998, Heskes proposed a bias-variance decomposition scheme for the KL divergence~\cite{heskes1998bias}, which would read in our problem as
\begin{equation}
\left\langle D_{\mathrm{KL}}(P^0_V\|Q^m_V)\right\rangle_m = \underbrace{D_{\mathrm{KL}}(P^0_V\|\bar{Q}_V)}_{\equiv \mathrm{Bias}}+\underbrace{\left\langle D_{\mathrm{KL}}(\bar{Q}_V\|Q^m_V)\right\rangle_m}_{\equiv \mathrm{Variance}},
\end{equation}
where
\begin{equation}
\bar{Q}_V(\mathbf{v}) \equiv \frac{1}{\mathcal{Z}}\exp\,\langle \log Q^m_V(\mathbf{v}) \rangle_m
\end{equation}
is the {\em mean distribution} for the suitable normalization constant $\mathcal{Z}$. This generalizes the original bias-variance decomposition originally proposed for the MSE in the following sense:
\begin{enumerate}
\item The variance does not depend on $P^0_V$ directly. Also it is nonnegative and equal to zero if and only if $Q^m_V$ are always identical.
\item The bias depends on only $P^0_V$ and the ``average model'' $\bar{Q}_V$, which is defined as the minimizer of the variance.
\end{enumerate}

Note that this scheme is related to our ME-DE decomposition scheme by
\begin{eqnarray}
	\mathrm{ME} &= \mathrm{Bias} - \sum_\mathbf{v} P^0_V(\mathbf{v}) \log \frac{Q^0_V(\mathbf{v})}{\bar{Q}_V(\mathbf{v})}, \\
	\mathrm{DE} &= \mathrm{Variance} + \sum_\mathbf{v} P^0_V(\mathbf{v}) \log \frac{Q^0_V(\mathbf{v})}{\bar{Q}_V(\mathbf{v})}.
\end{eqnarray}
Since $\sum_\mathbf{v}P^0_V(\mathbf{v}) \log \left[Q^0_V(\mathbf{v})/\bar{Q}_V(\mathbf{v})\right]$ is not sign definite (note that $\bar{Q}_V$ might be a distribution that cannot be generated by an RBM, so it can be even ``better'' than the optimal RBM-generated distribution $Q^0_V$), there are no definite inequalities between these error components.

We reexamine the effects of $N_H$ shown in Fig.~\ref{fig:tradeoff} using the Heskes decomposition scheme. As shown in Fig.~\ref{fig:heskes}, while the variance monotonically increases with $N_H$, the bias exhibits a nonmonotonic behavior as $N_H$ is increased. It seems that, in this case, the bias also contains significant contributions from the sampling fluctuations which were captured by the DE of our decomposition scheme. This might be due to the training outcome $Q^m_V$ having a skewed distribution around the optimal distribution $Q^0_V$. Thus, to describe the tradeoff behavior of the GE, the decomposition into the ME and the DE seems more appropriate than the Heskes bias-variance decomposition.

\section{Summary and outlook} \label{sec:summary}

\begin{table}
\begin{center}
\begin{tabular}[\textwidth]{c|c|c}
	& Supervised learning (FNN) & Unsupervised learning (RBM) \\ \hline
Generalization error & Mean-squared error & KL divergence\\
Limitation of the model	& Bias & Model error \\
Performance variability	& Variance & Data error \\
What it learns	& Functional relationship & Distribution \\
What it overfits & Noise & Sampling error \\
\end{tabular}
\caption{\label{tab} A comparison between the GE tradeoff behaviors in supervised (feed-forward neural network; FNN) and unsupervised learning}
\end{center}
\end{table}

In this study, we proposed that the generalization error (GE) in unsupervised learning can be decomposed into two components, namely the model error (ME) and the data error (DE). To examine how these quantities behave as the data and the model complexities are varied, we trained the RBMs with various numbers of hidden nodes to generate the steady-state configurations of the 2-d Ising model at a given temperature and the TASEP with given entry and exit rates, whose statistics are exactly known. For both models, we observed that the DE tends to decrease as the inverse volume of the training dataset, verifying that our decomposition properly distinguishes between the two sources of the GE, namely the inadequacies of the model (ME) and those of the training data (DE). Moreover, we found that the ME (DE) decreases (increases) monotonically as the number of hidden nodes is increased, leading to the tradeoff behavior of the GE. The GE minimum occurs at a higher number of hidden nodes when there exist stronger correlations between different parts of the system. This is analogous to the bias-variance tradeoff in supervised learning---too simple machines fail to capture the essential features of the data, while too complex machines fail to filter out the noise.

Thus, our study clarifies the nature of the GE tradeoff in unsupervised learning. Various theoretical studies are possible from here. For example, while our study has reported only numerical results for the RBMs, it would be interesting to think of an analytically tractable example of unsupervised learning for which the location of the GE minimum can be explicitly connected to the statistical properties of the training data. Also notable is the abrupt change in the training dynamics at the onset of the overfitting regime shown in Fig.~\ref{fig:dynamics}. It would be worthwhile to check whether the RBM crosses any phase boundary as its dynamical behavior changes. The issues of (i) how regularization suppresses overfitting and affects the GE minimum, (ii) how the hidden layer differently encodes information in the underparametrized and the overparmetrized regimes~\cite{song2018resolution,harsh2020place}, and (iii) whether the double descent of the GE observed in supervised learning~\cite{belkin2019reconciling,spigler2019jamming,nakkiran2021deep, rocks2022memorizing} is also possible in unsupervised learning would be also interesting to investigate.

\begin{figure}
\includegraphics[width=\columnwidth]{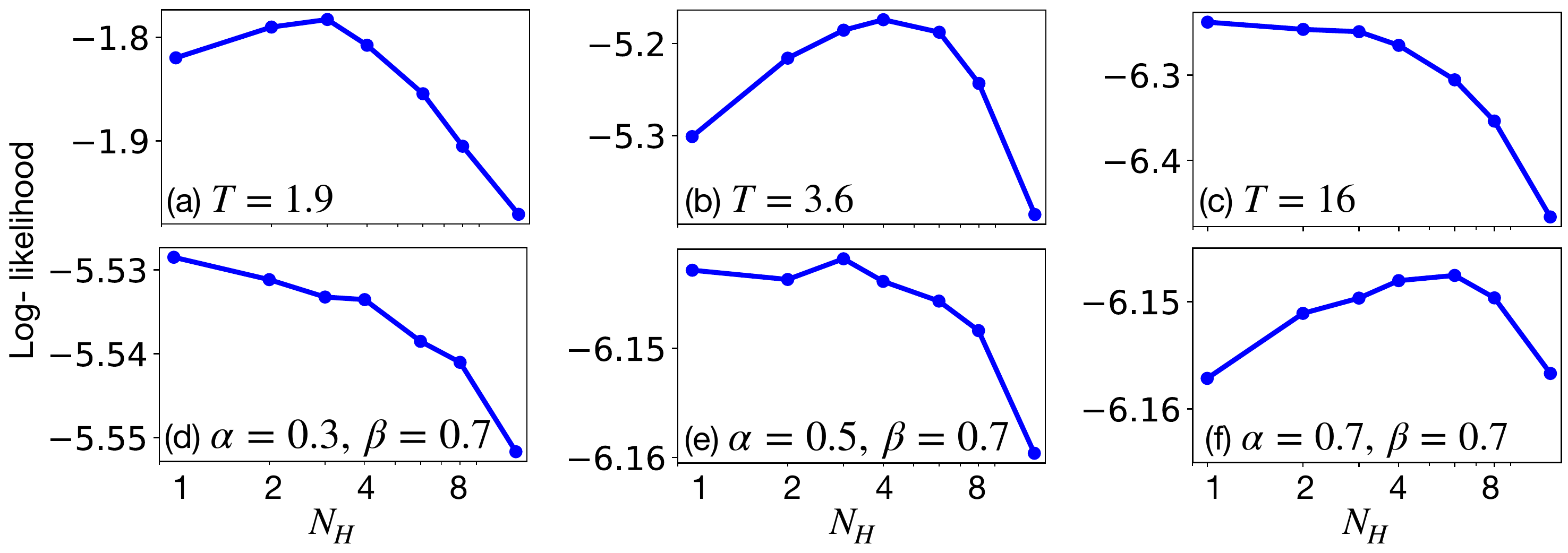}
\caption{\label{fig:likelihood} Dependence of the log-likelihood on the number of hidden nodes in the RBM. We show results for the 2-d Ising model at (a) $T = 1.9$ (ferromagnetic), (b) $T = 3.6$ (critical), (c) $T = 16$ (paramagnetic) and for the TASEP with open boundaries at (d) $\alpha = 0.3$, $\beta = 0.7$ (low density), (e) $\alpha = 0.5$, $\beta = 0.7$ (phase boundary), (f) $\alpha = 0.7$, $\beta = 0.7$ (maximal current). The error bars are explicitly shown for the Ising model, while they are much smaller than the symbol size for the TASEP.}
\end{figure}

On the practical side, we note that the GE defined as the KL divergence in Eq.~\eref{eq:ge_def} is not easy to calculate as $P^0_V$ is unknown in practice. Instead, one can focus on the tradeoff behavior of the log-likelihood
\begin{equation}
\mathcal{L} \equiv \left\langle\frac{1}{M}\sum_{i=1}^M \log Q^\mathrm{m}_V(\mathbf{v}_i)\right\rangle_\mathrm{m},
\end{equation}
whose value is obtained using a sampled test dataset $\{\mathbf{v}_1,\ldots,\mathbf{v}_M\}$. Unlike the GE, $\mathcal{L}$ is not bounded by zero, but the the nonmonotonic behavior of the GE as a function of $N_H$ will manifest itself as the inverted nonmonotonic behavior of $\mathcal{L}$, as illustrated for the 2-d Ising model and the TASEP in Fig.~\ref{fig:likelihood}. It would be interesting to check whether the same behavior is observed for more diverse range of examples.


\section*{Acknowledgments}

This research has been supported by the POSCO Science Fellowship of POSCO TJ Park Foundation (GK, HL, and YB). It has also been supported in part by the Creative-Pioneering Researchers Program through Seoul National University, the National Research Foundation of Korea (NRF) grant (Grant No. 2022R1A2C1006871) (JJ).

\section*{References}
\bibliographystyle{iopart-num-mod}
\bibliography{bbv_ref}

\end{document}